\documentclass[compsoc,conference,a4paper,10pt,times]{IEEEtran}
\pdfoutput=1 

\usepackage[backend=biber,
    bibstyle=ieee,
    sortcites=true,
    maxnames=2,
    maxbibnames=12,
    minnames=1,
    defernumbers=false,
    uniquelist=false,
]{biblatex}

\usepackage[colorlinks=true,urlcolor=black]{hyperref}
\def\BibTeX{{\rm B\kern-.05em{\sc i\kern-.025em b}\kern-.08em
    T\kern-.1667em\lower.7ex\hbox{E}\kern-.125emX}}

\usepackage{booktabs}
\usepackage{tabularx}
\usepackage{multirow}
\usepackage{rotating}
\usepackage{csquotes}
\usepackage{tikz}
\usepackage{pgfplots}
\usetikzlibrary{
        shapes,
        decorations
    }

\usepackage{amsmath,amssymb,amsfonts}
\usepackage{cleveref}
\crefname{figure}{Fig.}{Figures}
\crefname{table}{\tablename}{Tables}
\usepackage{algorithmic}
\usepackage{graphicx}
\usepackage{textcomp}
\usepackage{xcolor}
\usepackage{xspace}
\usepackage{subcaption}
\usepackage{interval}
\usepackage[inline]{enumitem}
\newcommand*{\eg}{e.g.\@\xspace}
\newcommand*{\ie}{i.e.\@\xspace}
\newcommand*{\ourmodel}{JOIST\xspace}

\definecolor{sron0}{HTML}{332288}
\definecolor{sron1}{HTML}{88CCEE}
\definecolor{sron2}{HTML}{117733}
\definecolor{sron3}{HTML}{DDCC77}
\definecolor{sron4}{HTML}{CC6677}
\definecolor{sron5}{HTML}{AA4499}

\begin{document}


%\title{A Predictive Model for the Zcash Verification Delay}

% Titelvorschläge:
%\title{Estimating the Zcash Verification Delay}
%\title{A Model for the Zcash Block Verification Time}
%\title{Modelling Zcash Block Verification Time}
\title{Modeling the Block Verification Time of Zcash\vspace{-0.7em}}
%\title{Analysing Zcash Block Verification Time}
%\title{Analysing Block Verification in Zcash}

\author{\IEEEauthorblockN{Fabian Stiehle}
 \IEEEauthorblockA{%\textit{Distributed Security Infrastructures} \\
 \textit{Technische Universität Berlin}\\
 % Berlin, Germany \\
 stiehle@campus.tu-berlin.de\vspace{-1ex}}
 \and
 \IEEEauthorblockN{Erik Daniel}
 \IEEEauthorblockA{%\textit{Distributed Security Infrastructures} \\
 \textit{Technische Universität Berlin}\\
 %Berlin, Germany \\
 erik.daniel@tu-berlin.de\vspace{-1ex}}
 \and
 \IEEEauthorblockN{Florian Tschorsch}
 \IEEEauthorblockA{%\textit{Distributed Security Infrastructures} \\
 \textit{Technische Universität Berlin}\\
 %Berlin, Germany \\
 florian.tschorsch@tu-berlin.de\vspace{-1ex}}
 }

\maketitle

\begin{abstract}
% Long block propagation times can lead to more stale blocks and
% respectively increase the occurence of forks in blockchain networks.
An important aspect of the propagation delay in blockchain networks is the block verification time,
which is also responsible for the so-called verifier's dilemma.
Models for the block verification time can
help to understand and improve the verification process.
Moreover, modeling the verification time is necessary for blockchain network simulations.
In this paper, we present \ourmodel, a new model for the block verification time of Zcash.
We identify computationally complex operations in the verification process of Zcash, and
derive our model based on characteristic transaction features.
We evaluate \ourmodel and show that the model
is consistently more accurate than existing models,
which consider the block size only.
\end{abstract}


\section{Introduction}\label{sec:intro}
The independent verification of blocks is a core principle of a
permissionless blockchain.
While the verification of a block header is simple,
the verification of transactions in a block can become much more time consuming.
The large number of transactions and the complexity of transactions
drives the required verification time.
This leads to a fundamental design flaw,
which is captured by the so-called verifier's dilemma~\cite{luu2015demystifying}.
It describes the conflict of a miner
between verifying a block for the \enquote{common good}
and skipping expensive verifications altogether.
In either way, the consequences are unfavorable
as the network and in particular miners become susceptible to attacks.

Understanding the verification of blocks and transactions
is therefore crucial to analyze and improve the mechanics of blockchain networks.
For example, the verification time plays a role in
the block distribution~\cite{motlagh2020analytical,rohrer2019kadcast,shahsavari2020theoretical},
blockchain simulations~\cite{gervais2016security},
as well as topology inference~\cite{daniel19mapz}.
%FLO: Hier nochmal ein Security-Argument?
Today's models estimating the block verification time
mainly rely only on the block size as a rough approximation.
The verification time, however, depends on the complexity of included transactions
and therefore can differ between blocks of the same size.

Zcash~\cite{zcash2020prot} is a prime example for complex transaction scripting,
where an accurate model of the verification time is highly relevant.
Zcash uses various transaction types and zero-knowledge proofs,
which are more complex than Bitcoin's transaction scripting~\cite{nakamoto2009bitcoin}.
While in Bitcoin the transfer time seems to be the dominating factor
of block propagation~\cite{kanda2019estimation},
the verification time in Zcash can be expected to have a higher impact
on system properties
as the network is smaller and block propagation delays are shorter~\cite{daniel19mapz}.
Thus, more sophisticated models are required.

In this paper, we present \emph{\ourmodel},
a novel---yet deliberately simple---model for the block verification time of Zcash.
Our model considers different transaction features %included in a block
to predict the block verification time.
We evaluate \ourmodel based on a benchmark of real-world Zcash block verification time samples.
Our model has a lower error compared to block size-based models,
yielding a mean absolute error of $3\,ms$ compared to approximately $11\,ms$.
% and a $R^{2}$ score as high as $0.91$ compared to $0.25$.
In addition, \ourmodel captures extreme values better,
which have a higher impact on the overall block propagation.
Our results also provide general insights,
\ie, varying transaction complexity of transactions not specific to Zcash
render block size-based models inaccurate even beyond Zcash.

Our main contributions are (i) a model for the block verification time of Zcash
based on transaction features and
(ii) a benchmark of the Zcash client's block verification
process and its employed cryptographic primitives to support and evaluate our model.

The remainder is structured as follows.
After reviewing related work in~\Cref{sec:relwork},
%Then we give a general overview of Zcash and
%the block and transaction verification process in~\Cref{sec:zcash}.
we introduce \ourmodel in~\Cref{sec:model}.
In~\Cref{sec:eval}, we show the benchmark and evaluation of our model.
% We discuss in~\Cref{sec:discussion} design decisions and weaknesses of our model
% as well as possible problems for adopting the model to other cryptocurrencies.
\Cref{sec:conclusion} concludes the paper.

\section{Related Work}\label{sec:relwork}
Block propagation is an important aspect
of analyzing and modeling blockchain networks.
% \citeauthor{decker2013propagation}~\cite{decker2013propagation} investigate the
% propagation delay in an early version of Bitcoin.
% They argue that the verification delay is
% a major factor for the overall propagation delay and
% show that an increased propagation delay increases the likelihood of blockchain forks.
Some approaches try to reduce the impact of block verification~\cite{das2020better,liu2019reducing,eberhardt2018zokrates,bip152},
which is orthogonal to our work.

Blockchain simulation models need to consider verification time
or need to implement the verification procedure.
Simulations and analytical models abstract the Bitcoin
logic~\cite{gervais2016security,rohrer2019kadcast,motlagh2020analytical,shahsavari2020theoretical}.
These approaches assume a linear block size-based verification time.
Block verification can also be a measurement based delay~\cite{miller2015shadow,faria2019blocksim}.

Research directly related to Zcash's verification time also exists:
The Zerocash paper~\cite{ben2014zerocash} approximates verification time
for a transactions with a constant of $10\,ms$.
Another block size-based model fitted for Zcash
considers verification time differences of different nodes~\cite{daniel19mapz}.
To the best of our knowledge, we present the first block verification
model independent of block size for Zcash.
%Contrary to previous approaches, we only model the block verification process.
% While our model is independent of the block size,
% we use the linear block size-based model as comparison.

%\input{zcash}
\section{\ourmodel Verification Model}\label{sec:model}
We analyzed the Zcash code and identified relevant predictors
for the block verification time.
% Im using predictor when its a feature used in the model, and feature as equivalent to "column" in our data
While verifying the block header can be considered constant,
the verification steps for transactions differ greatly
based on the set of transaction types.
More specifically, we consider the number of
\textbf{J}oinSplits,
\textbf{O}utput descriptions,
\textbf{I}nputs, and
\textbf{S}pend descriptions in
\textbf{T}ransactions, \ie, \emph{\ourmodel}.

\subsection{Transaction Complexity}
Zcash uses shielded and transparent transactions.
While transparent transactions are comparable to Bitcoin transactions,
shielded transactions use zero knowledge proofs to hide sender and receiver information
as well as the amount of Zcash coins (ZEC) in a transaction.
Since the Sapling network upgrade shielded transactions contain Spend and Output descriptions instead of so-called JoinSplit descriptions.
In the following, we identify and enumerate expensive computations
in the transaction and block verification process of Zcash.
At the same time, we strive to find simple features to predict the complexity.

For transparent transactions, we abstract from the scripting engine
and assume P2PKH scripts, which require one signature check.
Therefore, we assume verification time scales linearly with
the number of transparent inputs.
%Please note that since we count inputs,
%it already leads to a more fine-granular level than most block size-based models.

For shielded transactions, each Spend, Output,
and JoinSplit description incurs a separate zk-SNARK proof.
Verifying a single zk-SNARK proof can be computed in~$\mathcal{O}(1)$.
We therefore assume a linear relationship between the verification time of zk-SNARK proofs
and the number of transparent inputs, Spend and Output descriptions, and JoinSplits.

In addition, all transaction types require repeated calculations of transaction hashes, \eg, to check signatures.
One could conclude that the verification time scales linearly
with the number of verification steps.
% Moreover, hash operations would incur $\mathcal{O}(n^2)$,
% as the hash depends on the transaction size, which would in turn,
% partly depend on the amount of (hash) operations included in the transaction.
% Since the Overwinter upgrade, however,
Zcash, however, allows re-use of transaction hashes
by caching them.\footnote{See
\textit{ZIP: 143, Transaction Signature Validation for Overwinter}
\url{https://zips.z.cash/zip-0143} and
\textit{ZIP: 243, Transaction Signature Validation for Sapling}
\url{https://zips.z.cash/zip-0243} [Both Accessed: 02-Sep-2020].}
Therefore, they do not have to be re-computed
for multiple verification steps across one transaction,
% improving signature verification to $\mathcal{O}(n)$ and making it largely independent from transaction size.
yielding a constant complexity per transaction.
We similarly model the hash calculations on the block level
as constant operation.

% In shielded transaction, the transaction hash cache pool
% ensures signature verification is only done once per transaction.
% JoinSplit specific signatures are only verified,
% if a JoinSplit description is present.
% % We will discuss the likelihood of
% % JoinSplit descriptions occuring in a block momentarily.
% Each Spend description requires verifying a key and signature.

% Due to the transaction hash cache,
% we model hash operations as a constant.
% We extend this abstraction to the block level
% and model the overhead resulting
% from hash operations as a constant per block.

Since transparent outputs are referenced by inputs
and do not trigger complex computations on their own,
we can ignore them in our model.

\subsection{Statistical Analysis}
We analyze the statistical correlation between
different transaction properties and verification time.
%Although outputs are not directly involved in the verification process,
%we include the number of outputs for verification.
Therefore, we calculate Pearson's correlation coefficient~$r$
for each selected feature in relation to the block verification time~(cf. \Cref{tab:pcc_all}).
As a sample we use Zcash blocks from block height 715,578 to block height 915,578,
which were the most recent blocks at the beginning of our work.
The verification time is determined in a benchmark, described in detail in \Cref{sec:eval}.
For each observation $i$ in our sample,
\ie, for each observed block height,
let~$x_i$ be our feature and let~$t_i$ be the verification time.
Then
\begin{align}
r = \frac{\sum^n_{i=1}(x_i-\bar{x})(t_i-\bar{t})}
{\sqrt{\sum^n_{i=1}(x_i-\bar{x})^2}\sqrt{\sum^n_{i=1}(t_i-\bar{t})^2}},
\end{align}
where $n$ is the sample size, $\bar{x}$ is the mean over all feature observations,
%\ie, $\bar{x} = \frac{1}{n} \sum_{i=1}^{n} x_i.$
and $\bar{t}$ is the mean block verification time.

We show the results for each selected feature in \Cref{tab:pcc_all}.
The results show that the number of JoinSplits, Output descriptions, transparent inputs, and Spend descriptions of the transactions
exhibit a linear relationship with respect to the verification time.
% FS: Changed "significantly", as p<0.1 proofs the significance also for outputs
Outputs have only a small effect on the verification process,
since they are not involved in the process.

% \begin{figure}
%   \centering
%   \includegraphics[width=\columnwidth]{figures/top.pdf}
%   \caption{Ratio of transparent inputs, Spend and Output descriptions,
%   and JoinSplit descriptions per block (200,000 blocks).}
%   \label{fig:topologie}
% \end{figure}
%
% Notably, we can see a strong correlation for the number of JoinSplits.
% This is surprising as JoinSplits are a legacy transaction type,
% which was superseded by Sapling's Spend and Output descriptions.
The analysis further shows that JoinSplits, although legacy, cannot be ignored.
We identified many blocks still containing JoinSplits.
Once a JoinSplit is included, verification time increases significantly
as it often includes the verification of two zk-SNARK proofs.

%In fact, as we show in \cref{fig:topologie},
%the number of JoinSplits remains largely stable over the last 200,000 blocks in our sample.
% (block height 715,578--915,578).
%Here, we consider a block as a set of transparent inputs,
%Spend, Output, and JoinSplit descriptions.
%We find that blocks consist on average of 90\,\% transparent inputs,
%while Spend an Output descriptions make up 9\,\%
%and JoinSplit descriptions remain stable at around 1\,\% per block.
%While the share of JoinSplits seems to be negligible,
%once a JoinSplit is included, the verification time increases significantly
%as it typically includes the verification of up to two zk-SNARK proofs.
%We found one JoinSplit description incurs an average verification time of $5993\,\mu s$,
%while a transparent input incurs $0.2\,\mu s$.
% Moreover, in our 200,000 block sample we encounter a block
% containing at least one JoinSplit transaction with probability $0.06$.
%While legacy, JoinSplits cannot be ignored when modeling the verification time.

\subsection{Model}
We consider the number of
JoinSplit~$n_{j}$, Output descriptions~$n_{o}$, transparent Inputs~$n_{i}$, and
Spend descriptions~$n_{s}$ of all transactions in the block
as the predictors of \ourmodel.
We derive our model to estimate the verification time~$\hat{t}$ of a given block as
\begin{align}
\hat{t} = \beta_{j}n_{j} + \beta_{o} n_{o} + \beta_{i}n_{i} + \beta_sn_{s} + k.
\end{align}
%
%In the model, $n_{v}$ is the number of transparent inputs,
%$n_{s}$ is the number of Spend descriptions,
%$n_{o}$ the number of Output descriptions,
%and $n_{j}$ the number of JoinSplits.
The constant $k$ accounts for verification overhead of block header,
hashing transactions, and possible I/O operations.

The values $\beta_{j}$, $\beta_o$, $\beta_{i}$, and $\beta_s$ are constant coefficients used as weights.
The coefficients for the model can be found by performing a benchmark on a local client.
It is certainly possible to parametrize the model
by using the mean or median time of the respective verification operations.
% (depending on the quality of the sample distribution).
For improved accuracy, however, coefficients should be discovered by linear regression.
For our evaluation, we use ordinary least squares to parametrize \ourmodel.

\begin{table}
  \scriptsize
  \centering
  \caption{$r$ for the verification time of 200,000 blocks}
  \begin{tabularx}{\columnwidth}{Xll}
    \toprule
    \textbf{Feature (number of)} & \textbf{$r$} & \textbf{$p$-Value}\\
    \midrule
    Transparent inputs & 0.11 & $< 0.01$\\
    Transparent outputs & 0.04 & $< 0.01$\\
    Spend descriptions & 0.38 & $< 0.01$\\
    Output descriptions & 0.53 & $< 0.01$\\
    JoinSplit descriptions & 0.60 & $< 0.01$\\
    \bottomrule
  \end{tabularx}
  \label{tab:pcc_all}
\end{table}

\section{Evaluation}\label{sec:eval}
In this section, we evaluate \ourmodel and compare it to
block size-based models used in literature.
In order to parametrize the models,
we utilize two benchmarks on two different systems.

\begin{table}
\scriptsize
\centering
\caption{Benchmark system specification}
\begin{tabularx}{\columnwidth}{lXX}
\toprule
 & \textbf{HDD} & \textbf{SSD} \\
\midrule
Cores & 2 & 6 \\
Processor & Intel Core i5 & AMD Ryzen 5 2600X\\
Clock rate & $2.6\,GHz$ & $3.60\,GHz$ \\
RAM, SWAP & $5\,GB$, $2\,GB$ & $12\,GB$, $2\,GB$ \\
Disk & WD Elements\newline USB 3.0 & Samsung\newline SSD 860 Evo \\
Disk I/O & 100\,MB/s & 540\,MB/s \\
\bottomrule
\end{tabularx}
\label{tab:bench_spec}
\end{table}

\subsection{Benchmark Setup}\label{sec:benchmark_analysis}
We perform the benchmarks utilizing two different setups.
One benchmark is performed on a low performance system
with an HDD disk, limited RAM, and restricted CPU.
The other system is a high(er) performance, consumer-grade system
with an SSD disk, more RAM, and faster CPU.
The specific hardware configurations are shown in \cref{tab:bench_spec}.
For short, we refer to the two systems as HDD and SSD,
because we believe that the access times in general
will have a major impact on our results.
More specifically, we use the limited HDD system to show the influence of
disk access and slower CPU times on the overall verification process.
In contrast, we use the SSD system to simulate more realistic client hardware.

For our benchmark, we perform a black box-like benchmark
on both systems, capturing the complete block verification process.
The verification process amounts to the time the client processes
the \texttt{ProcessNewBlock} function during synchronization with the network.
For a fine grained analysis, we separately performed benchmarks of singular computational steps,
such as verifying singular inputs or spends, performed during verification.
Our evaluation is based on different sequences
starting with the latest block in our sample at block height $915,578$.
% taken from block height $695,581$ to height $915,578$.
For improved reproducibility,
we compiled and run the Zcash 3.0 client inside a Docker container.
The used code and data sets are publicly
available.\footnote{https://github.com/fstiehle/zcash-benchmark}

\subsection{Model validation}
\label{sec:eval_novel}
\newcommand{\tableSpace}{2pt}
\newcommand{\tableExtraRule}{.6pt}

In order to evaluate \ourmodel,
we consider the mean absolute error~(MAE)
as well as the error mean ratio~(EMR),
which we define for completeness in Appendix~\ref{sec:appendix_calc}.
%We define the error $e_i$ as the difference between the real verification time $t_i$ and the prediction $\hat{t}_i$,
%\ie, $e_i = t_i - \hat{t}_i.$
%Moreover, let $\bar{t}$ be the average verification time over our complete sample.
%\ie, $\bar{t} = \frac{1}{n} \sum_{i=1}^{n} t_i.$
%We accordingly define MAE and EMR as
% \begin{align}
% \bar{t}&= \frac{1}{n} \sum_{i=1}^{n} t_i, \\
% \text{MAE}&= \frac{\sum_{i=1}^{n} \left| e_i \right|}{n} \quad\text{and}\\
% \text{EMR}&= \frac{\text{MAE}}{\bar{t}}.
% \end{align}
In order to evaluate the general fit of our model,
we calculate the coefficient of determination $R^2$
and the adjusted $R^2$, denoted as $\bar{R}^2$.
Let $n$ be the sample size (number of blocks),
$\hat{t}_i$ be the predicted,
$t_i$ the observed verification time at block height $i\in n$,
and $\bar{t}$ be the average verification time over our complete sample.
Therefore, $R^2$ and $\bar{R}^2$ can be calculated as
\begin{align}
R^2&= 1 - \frac{\sum_{i=1}^{n} (t_i-\hat{t}_i)^2}{\sum_{i=1}^{n} (t_i - \bar{t})^2} \quad\text{and}\\
\bar{R}^2&= 1 - (1- R^2)\,\frac{n-1}{n-p-1},
\end{align}
where $p$ is the number of the model's predictors.

%Finally, we test the statistical signifcance of our predictors by calculating the %F-statistic ($F$) as follows.

%\begin{align}
%  F = \frac{R^2\,/\,(p-1)}{1 - R^2\,/\,(n-p)}
%\end{align}

For the model validation, we use a sample size of 15k and 220k blocks.
% We choose the last 15,000 blocks from block height 915,578 to avoid legacy
% transactions as much as possible and depict the current state of Zcash.
We then uniformly at random choose 5k blocks to fit the model
and use the remaining 10k blocks to predict the verification time.
For the 220k block sample, we analogously use 20k to fit the model
and use the remaining 200k blocks to predict verification times.
% The evaluation is therefore based on the prediction results.
We use ordinary least squares~(OLS) to discover coefficients for \ourmodel,
which are listed in Appendix~\ref{sec:appendix_parameter}.

% \begin{table}
% \small
% \centering
% \caption{$d_v$ model evaluation. We use 5,000 blocks for parametrization and predict 10,000 blocks.}
% \begin{tabularx}{\columnwidth}{Xrrrrr}
% \toprule
%  & \textbf{MAE} & \textbf{EMR} & \textbf{$R^2$} & \textbf{$\bar{R}^2$}\\
% \midrule
% HDD & 22,051$\,\mu s$ & 64\% & 0.16 & 0.16 & \\
% SSD & 3,109\,$\mu s$ & 23\% & 0.91 & 0.91 & \\
% \bottomrule
% \end{tabularx}
% \label{tab:dv_eval}
% \end{table}

In \cref{tab:evaluation}, we summarize the results for the HDD and SSD benchmark.
Notably for the SSD benchmark predicting 10k blocks,
\ourmodel was able to predict the block verification time
with $\text{MAE} = 3\,ms$ and $\text{EMR} = 23\,\%$.
Moreover, our model was able to explain most of the variance with $R^2=0.9$.
For the SSD benchmark predicting 200k blocks,
where approximately 10\,\% of blocks were used for parametrization,
\ourmodel performed expectedly less accurate.
However, \ourmodel sill achieves an $\text{MAE} = 7\,ms$, an $\text{EMR} = 43\,\%$, and $R^2=0.7$.
%which is superior when compared to block size-based models.

\begin{figure*}
  \center
  \begin{subfigure}[c]{.32\textwidth}
      \includegraphics[width=\textwidth]{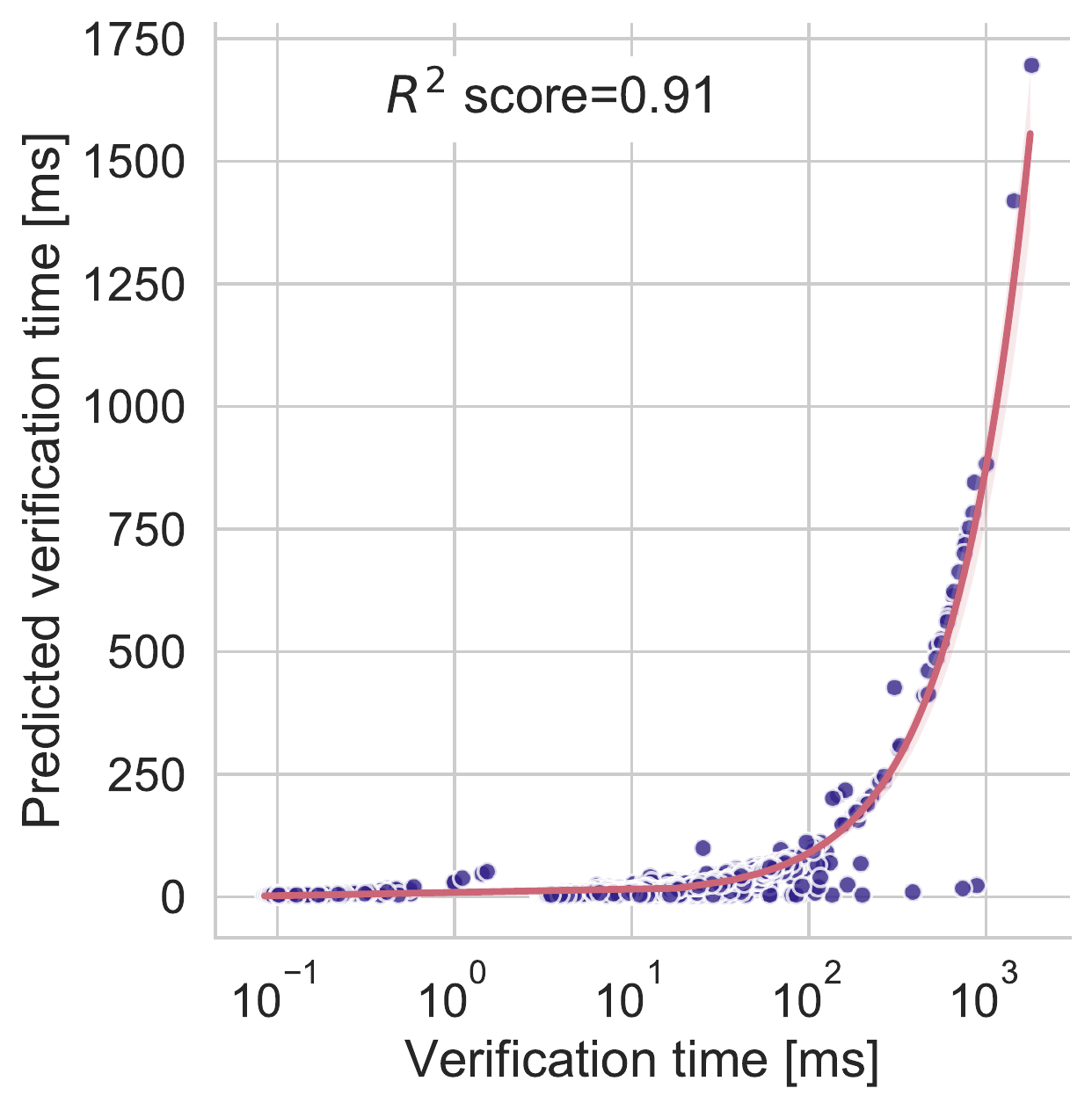}
      \caption{\ourmodel.}
      \label{fig:kv_r2}
  \end{subfigure}
  \begin{subfigure}[c]{.32\textwidth}
      \includegraphics[width=\textwidth]{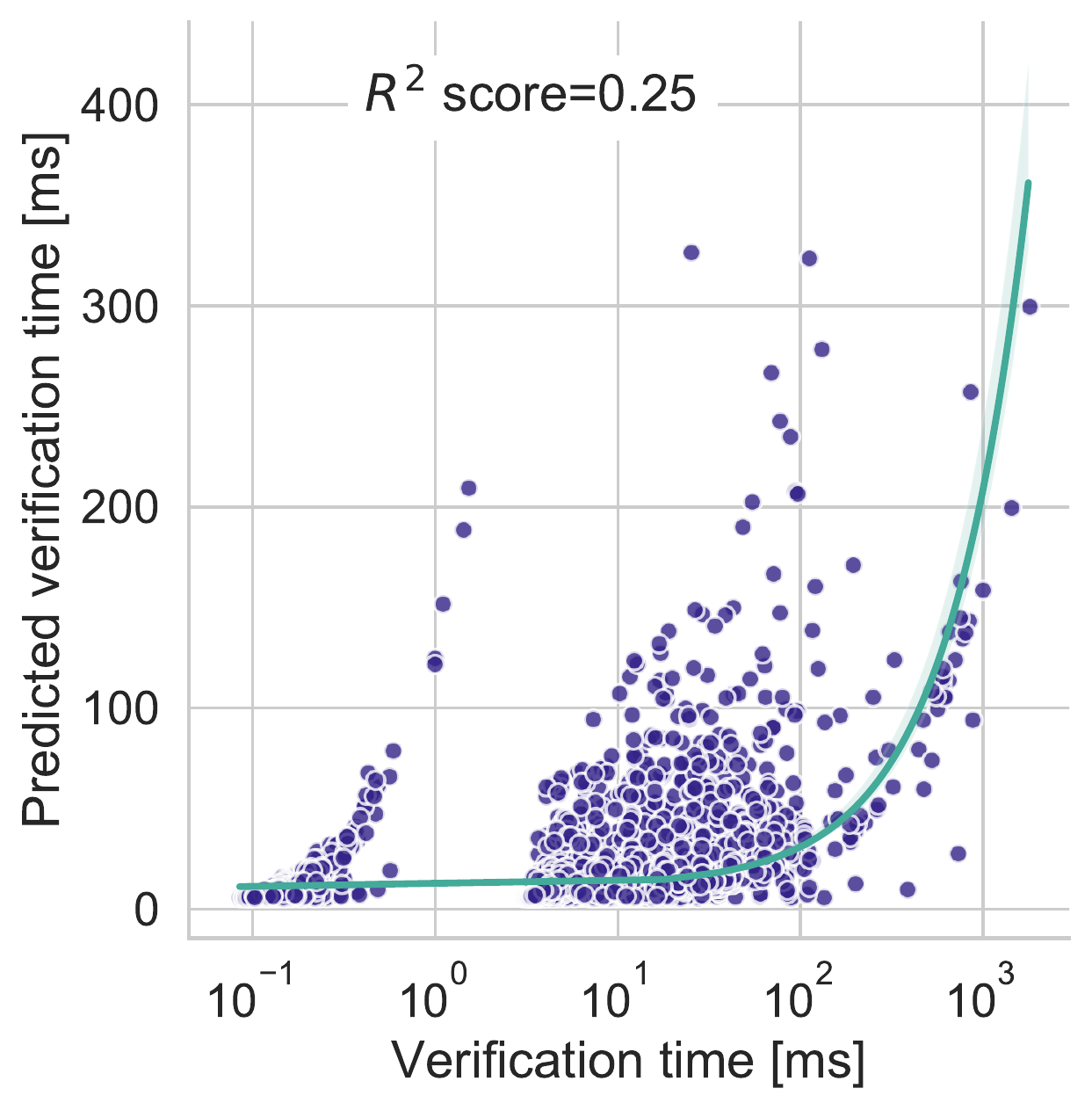}
      \caption{Block size.}
      \label{fig:kb_r2}
  \end{subfigure}
  \begin{subfigure}[c]{.32\textwidth}
    \includegraphics[width=\textwidth]{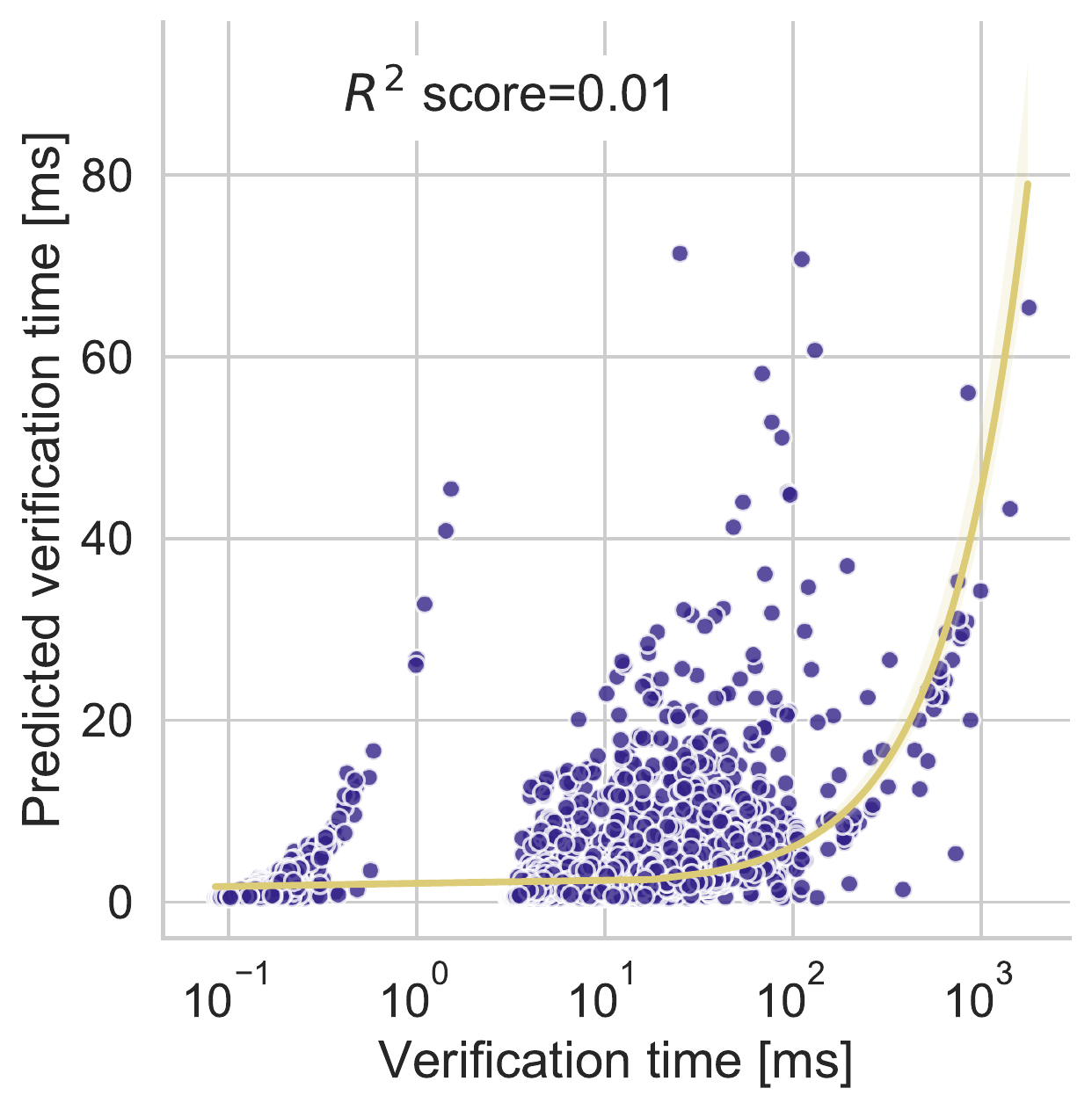}
    \caption{\citeauthor{gervais2016security}.}
    \label{fig:gb_r2}
\end{subfigure}
  \caption{Scatter plots of predicted and measured verification times
  with regression lines (10k blocks, SSD benchmark, log scaled x-axis).}
  \label{fig:r2}
\end{figure*}

In \cref{fig:kv_r2}, we compare estimated
with measured verification time for the SSD benchmark
(mind the log scale on the x axis).
While we see some variance for verification times between $10\,ms$ and $100\,ms$,
we observe a generally good fit.
Our model is even able to explain most of the outliers
with verification times over $250\,ms$.

% \begin{figure*}
%   \center
%   \includegraphics[width=\textwidth]{figures/KV_KB_Size.pdf}
%   \caption{Verification time in comparison to the block size
%   (10,000 blocks, y-axis zoomed for improved visualization).}
%   \label{fig:KB_KB_size}
% \end{figure*}

As expected, the HDD benchmark is more erratic.
The results yield an $\text{MAE} = 22\,ms$ and an $\text{EMR} = 64\,\%$ when predicting 10k blocks
and an $\text{MAE} = 30\,ms$ and an $\text{EMR} = 69\,\%$ when predicting 200k blocks.
%In \cref{fig:KB_KB_size}, we provide a visual impression of two experiment runs comparing HDD and SSD.
%More specifically, we show the predictions of our model and the real verification time observed in relation to the block size.
%We can clearly see how the slow disk access in the HDD benchmark
%leads to high peaks in the benchmark that our model is not able to explain.
For both benchmarks, $\bar{R}^2$ is very close to $R^2$
which suggests that all our predictors actually contribute to the prediction.

\subsection{Model Comparison}\label{sec:eval_compare}
Previous models~\cite{gervais2016security,rohrer2019kadcast,daniel19mapz,shahsavari2020theoretical,motlagh2020analytical}
assume a linear relationship between the block size and the block verification time.
In order to compare this approach to our model,
we introduce a generic block size-based model,
predicting the verification time~$\hat{t}$ of a block as
\begin{equation}
\hat{t} = \beta s_{b} + k,
\end{equation}
where $s_b$ is the block size, $\beta$ a constant weighting coefficient,
and $k$ a constant to capture disturbances.
In the following, we refer to this model as Block size model.

In addition, we use the model by \citeauthor{gervais2016security}~\cite{gervais2016security},
which is a block size-based model, assuming a mean block size of~$458,263\,B$ and mean validation time of~$0.174\,s$.
While it is designed for Bitcoin,
we include it in our evaluation due to the prevalent number of transparent transactions in Zcash,
which are identical to Bitcoin.
An overview over the distribution of different transaction types is provided in Appendix~\ref{sec:appendix_tx}.

% \begin{figure*}
%   \centering
%   \includegraphics[width=0.7\textwidth]{figures/KB_SIZE.png}
%   \caption{$d_{b}$ and the verification delay of 200,000 observed blocks during our benchmark on the HDD and SSD system.}
%   \label{fig:dv_size}
% \end{figure*}

\begin{table}
\scriptsize
\centering
\caption{Model comparison summary}
\label{tab:evaluation}
\begin{tabularx}{\columnwidth}{lXrrrrr}
\toprule
& \textbf{Model} & \textbf{$n$} & \textbf{MAE} & \textbf{EMR} & \textbf{$R^2$} & \textbf{$\bar{R}^2$} \\
\midrule
\multirow{6}{*}[-3pt]{\begin{sideways}SSD Benchmark\end{sideways}}
    & \ourmodel & 10k & 3.11$\,\text{ms}$ & 23\,\% & 0.91 & 0.91\\
    & & 200k & 7.11$\,\text{ms}$ & 43\,\% & 0.72 & 0.72\\
    \cmidrule(l){2-7}
    & Block size & 10k & 10.60$\,\text{ms}$ & 78\,\% & 0.25 & 0.25\\
    & & 200k & 15.13$\,\text{ms}$ & 92\,\% & 0.14 & 0.14\\
    \cmidrule(l){2-7}
    & \citeauthor{gervais2016security} & 10k & 11.66$\,\text{ms}$ & 86\,\% & 0.01 & 0.01 \\
    & & 200k & 14.60$\,\text{ms}$ & 89\,\% & 0.04 & 0.04 \\
\midrule
\multirow{6}{*}[-3pt]{\begin{sideways}HDD Benchmark\end{sideways}}
    & \ourmodel & 10k & 22.05$\,\text{ms}$ & 64\,\% & 0.16 & 0.16\\
    & & 200k & 30.03$\,\text{ms}$ & 69\,\% & 0.40 & 0.40 \\
    \cmidrule(l){2-7}
    & Block size & 10k & 31.20$\,\text{ms}$ & 91\,\% & 0.09 & 0.09\\
    & & 200k & 42.46$\,\text{ms}$ & 98\,\% & 0.12 & 0.12\\
    \cmidrule(l){2-7}
    & \citeauthor{gervais2016security} & 10k & 32.48$\,\text{ms}$ & 95\,\% & -0.03 & -0.03 \\
    & & 200k & 41.25$\,\text{ms}$ & 95\,\% & -0.01 & -0.01 \\
\bottomrule
\end{tabularx}
\end{table}

For the comparison, we evaluate MAE, EMR, $R^2$, and $\bar{R}^2$.
We use a sample size of 15k and 220k blocks
and uniformly at random choose 5k and 20k blocks to calculate coefficients, respectively.
The remaining blocks, 10k and 200k blocks, are used to predict verification times.
We use OLS to calculate coefficients for both models, JOIST and block size-based.
In Appendix~\ref{sec:appendix_parameter}, we list all fitting parameters for all evaluated models and benchmarks.
A summary of our results is shown in \cref{tab:evaluation}.
%In the following, we will present and discuss the results in detail.

% In \cref{fig:KB_KB_size},
% we compare \ourmodel with the Block size model as well as Gervais' model.
% We show the real verification time in relation with the block size for 10,000 predicted blocks
% for both the HDD and SSD benchmarks.
% The results highlight the non-linearity of the block size and the verification time.
% It shows that, while for the HDD system our model cannot explain the high peaks caused by slow disk access,
% it still performs better than the block size-based models.
% For the SSD benchmark, we can see how our model aligns closely
% with the general trend of the Benchmark.
% Even for outliers with block sizes larger than 100\,KiB,
% our model is able to perform accurate predictions,
% while the block size-based models either overestimate or underestimate the benchmark.

For the SSD benchmark predicting 10k blocks,
\ourmodel provides better estimations than both block sized-based models.
Our model was able to predict the verification time with an $\text{MAE}$ of $3\,ms$,
compared to approximately $11\,ms$.
Even when predicting 200k blocks,
using only 20k blocks for fitting the model,
our evaluation resulted in an $\text{MAE}$ of $7\,ms$, compared to $15\,ms$,
an EMR of $43\,\%$, compared to $89\,\%$,
and an $R^2=0.7$, compared to $0.14$.
% Into discussion: This underlines the predictive power of our model. (Low fitting size -> good results)

We compare the fit of the respective models in \cref{fig:r2},
where we show the estimated versus the measured verification time
and a fitted regression line.
We can observe a good fit for our model.
The block size-based models are unable to explain a number of variances,
particularly in the range between $0$ and $100\,ms$,
which contains most of our observed blocks.
The resulting low $R^2$ score of the block size-based models suggests
that these models make only slightly better predictions than the median.

In addition, \ourmodel performs well for extreme values,
which have a higher impact on the overall block distribution time
and are therefore important to simulate realistic block propagation,
for example.
While the highest prediction of \ourmodel was~$1,700\,ms$,
only four blocks~(0.02\,\%) exhibit a higher verification time.
In contrast, the maximal prediction of the Block size model was~$330\,ms$
and of \citeauthor{gervais2016security}~$71\,ms$.
That is, 68 blocks~(0.34\,\%) for the Block size model
and 2,849 blocks~(14\,\%) for \citeauthor{gervais2016security}
exhibit a higher measured verification time than the prediction.
As a result, \ourmodel exhibits a max error of~$840\,ms$,
the Block size model a max error of $1.5\,s$, and \citeauthor{gervais2016security} a max error of~$17\,s$.

\subsection{Discussion}\label{sec:discussion}
While we showed that our model predicts the block verification time of Zcash
more accurately than other models proposed in literature,
we observe that it performs much better for modern systems
with faster disk access, more RAM, and more CPU cores.
Low RAM forces the Zcash client to flush state to disk more frequently,
which leads to unpredictable access times.
%With the HDD benchmark, we deliberately forced this behavior.
%Accordingly, we can observe extreme peaks in the HDD benchmark (see \cref{fig:KB_KB_size})
%that are not visible in SSD benchmarks.
Similarly, the impact of multi-core, high(er) clock rate CPUs is visible.
Delays introduced by a single verification step, \eg, verifying a single input,
is a magnitude faster in the SSD benchmark~($\beta_{i}=61.411\,\mu s$)
when compared to the HDD benchmark~($\beta_{i}=246.312\,\mu s$).
By comparing both benchmarks and the performance of our model,
we can conclude that I/O operations are a negligible factor
during block verification on modern systems.

% TODO: How may blockchain forks influence the verification delay?

We additionally observe a high variance of verification time
for blocks containing transparent transactions only.
%Neither, \ourmodel, nor Block size, nor Gervais et al. were able to accurately capture this variance.
In such scenarios, our modeling approach is not able to accurately capture this variance.
%The number of inputs alone is not enough to derive an accurate model.
Likely, our abstraction of the scripting system,
\ie, our assumption to consider P2PKH scripts only,
led to this result.
Improving our model towards transparent transactions
will require a more in-depth analysis of the scripting system.
A major challenge would be to classify and identify relevant opcodes
with the highest impact on the verification time.
Extracting these features, including sample data,
would also be significantly more complex.
This direction, however, would increase our model's applicability
for other blockchains.

% While we focus on Zcash,
% we observed a high error for all models when predicting
% % transparent transactions ($\text{EMR} > 80\,\%$ and $R^2=0$~\cref{sec:appendix_ttx}).
% transparent transactions ($\text{EMR} > 80\,\%$ and $R^2=0$).
While we focus on Zcash,
we separately evaluated blocks with transparent transactions only.
We observed that all models perform similarly with a high error yielding
$\text{MAE} \approx 4\,\text{ms}$, $\text{EMR} > 80\,\%$, and $R^2=0$.
This indicates a general weakness.
That is, models using the block size as the only predictor
might be unable to predict block verification times accurately
for blockchains such as Bitcoin.
Therefore, not only blockchains with complex block verification procedures
such as Zcash or Ethereum~\cite{wood2014ethereum} require additional modeling efforts,
but also blockchains with a rather simple scripting system. %need further attention.

% TODO: Adoption to non UTXO permissionless Blockchains? Permissioned Blockchains?

In general, our model is a trade off between
abstraction, complexity, and transparency.
We deliberately use basic, easily acquirable features.
%Each transparent input, Spend, Output, and JoinSplit description in a block
Each feature is encoded in the Zcash RPC response and can be easily retrieved.

% while the impact of the individual features remain traceable.
% Other approaches, such as machine learning,
% often require expensive feature engineering.
% Nevertheless, we consider the application of explainable machine learning algorithms
% as future work.

\section{Conclusion}\label{sec:conclusion}
In this paper, we developed \ourmodel,
a new model based on transaction features
that provides more accurate results than
the prevalent block size-based models.
To this end, we extracted time-consuming primitives
from Zcash's transaction verification process.
\ourmodel is simple to parametrize and at the same time
able to predict the block verification time accurately.
% particularly for modern hardware.
% Our results suggest an average improvement of more than a factor of three.
% However, all models, including \ourmodel, are not able to explain the variance
% in the verification time of transparent transactions.
% This observation suggests that UTXO-based blockchains such as Bitcoin in general,
% and their scripting system in particular require additional attention in the future.
% For \ourmodel, we have discussed potential directions of improvement.

Overall, we believe that \ourmodel is a promising first step in modeling and predicting
the block verification time of blockchains,
which in turn can improve modeling and simulation efforts of the propagation delay
and blockchain systems in general.
Furthermore, the model could help miners to quantify verification risk
and therefore address the verifiers dilemma.

\printbibliography[heading=bibintoc]

% \newpage
\appendices
\section{Error Calculation}\label{sec:appendix_calc}
For completeness, we include the formulas for calculating MAE and EMR.
Let $n$ be the sample size, \ie, the number of blocks.
Let $\hat{t}_i$ be the predicted and $t_i$ the real observed verification time at block height $i\in n$.
We define the error $e_i$ as the difference between the real verification time $t_i$ and the prediction $\hat{t}_i$,
\ie, $e_i = t_i - \hat{t}_i.$
Moreover, let $\bar{t}$ be the average verification time over our complete sample,
\ie, $\bar{t} = \frac{1}{n} \sum_{i=1}^{n} t_i.$
We accordingly define MAE and EMR as
\begin{align}
\text{MAE}&= \frac{\sum_{i=1}^{n} \left| e_i \right|}{n} \quad\text{and}\\
\text{EMR}&= \frac{\text{MAE}}{\bar{t}}.
\end{align}

\section{Fitting Parameters}
\label{sec:appendix_parameter}
\vspace{-1ex}

% \begin{table}[h!]
% \scriptsize
% \centering
% \caption{Model fitting parameters}
% \begin{tabularx}{\columnwidth}{lrlX}
% \toprule
% \textbf{Model} & \textbf{Training} & \textbf{Benchmark} & \textbf{Parameters [$\mu s$]} \\
% \midrule
% \ourmodel & 5,000 & HDD &
% $\beta_{j} = 10999.119,\newline
% \beta_{o} = 9862.146, \newline
% \beta_{i} = 246.312,\newline
% \beta_{s} = 39760.496,\newline
% k = 13209.042$\\
% \cmidrule(l){3-4}
% & & SSD &
% $\beta_{j} = 5359.094,\newline
% \beta_{o} = 5726.675,\newline
% \beta_{i} = 61.411,\newline
% \beta_{s} = 16912.591,\newline
% k = 4468.949$ \\
% \cmidrule(l){2-4}
% & 20,000 & HDD &
% $\beta_{j} = 10784.519,\newline
% \beta_{o} = 12607.155,\newline
% \beta_{i} = 139.676,\newline
% \beta_{s} = 25227.674,\newline
% k = 21760.549$\\
% \cmidrule(l){3-4}
% & & SSD & 
% $\beta_{j} = 5349.659,\newline
% \beta_{o} = 5782.956,\newline
% \beta_{i} = 40.339,\newline
% \beta_{s} = 12067.658,\newline
% k = 5928.899$\\
% \midrule
% Block size & 5,000 & HDD & $\beta = 4.345\,\frac{1}{B},\newline k = 8784.760$ \\
% \cmidrule(l){3-4}
% & & SSD & $\beta = 1.717 \,\frac{1}{B},\newline k = 3584.715$ \\
% \cmidrule(l){2-4}
% & 20,000 & HDD & $\beta = 2.232\,\frac{1}{B},\newline k = 28445.511$ \\
% \cmidrule(l){3-4}
% & & SSD & $\beta = 0.910\,\frac{1}{B},\newline k = 9647.374$ \\
% \midrule
% \citeauthor{gervais2016security} & N/A & N/A & $\beta = 0.3796\,\frac{1}{B}$,\newline $k = 0$ \\
% \bottomrule
% \end{tabularx}
% \label{tab:fitting}
% \end{table}

\begin{table}[h!]
\scriptsize
\centering
\caption{Model fitting parameters}
\begin{tabularx}{\columnwidth}{llX}
\toprule
\textbf{Model} & \textbf{Benchmark} & \textbf{Parameters [$\mu s$]} \\
\midrule
\ourmodel & 5k, HDD &
$\beta_{j} = 10999.119,
\beta_{o} = 9862.146, \newline
\beta_{i} = 246.312,
\beta_{s} = 39760.496,\newline
k = 13209.042$\\
% \cmidrule(l){2-4}
& 5k, SSD &
$\beta_{j} = 5359.094,
\beta_{o} = 5726.675,\newline
\beta_{i} = 61.411,
\beta_{s} = 16912.591,\newline
k = 4468.949$ \\
% \cmidrule(l){2-4}
& 20k, HDD &
$\beta_{j} = 10784.519,
\beta_{o} = 12607.155,\newline
\beta_{i} = 139.676,
\beta_{s} = 25227.674,\newline
k = 21760.549$\\
% \cmidrule(l){3-4}
& 20k, SSD &
$\beta_{j} = 5349.659,
\beta_{o} = 5782.956,\newline
\beta_{i} = 40.339,
\beta_{s} = 12067.658,\newline
k = 5928.899$\\
\midrule
Block size & 5k, HDD & $\beta = 4.345\,\frac{1}{B}, k = 8784.760$ \\
% \cmidrule(l){3-4}
& 5k, SSD & $\beta = 1.717 \,\frac{1}{B}, k = 3584.715$ \\
% \cmidrule(l){2-4}
& 20k, HDD & $\beta = 2.232\,\frac{1}{B}, k = 28445.511$ \\
% \cmidrule(l){3-4}
& 20k, SSD & $\beta = 0.910\,\frac{1}{B}, k = 9647.374$ \\
\midrule
\citeauthor{gervais2016security} & N/A & $\beta = 0.3796\,\frac{1}{B}$, $k = 0$ \\
\bottomrule
\end{tabularx}
\label{tab:fitting}
\end{table}

\section{Transaction Types}\label{sec:appendix_tx}
\begin{figure}[h!]
   \centering
   \includegraphics[width=\columnwidth]{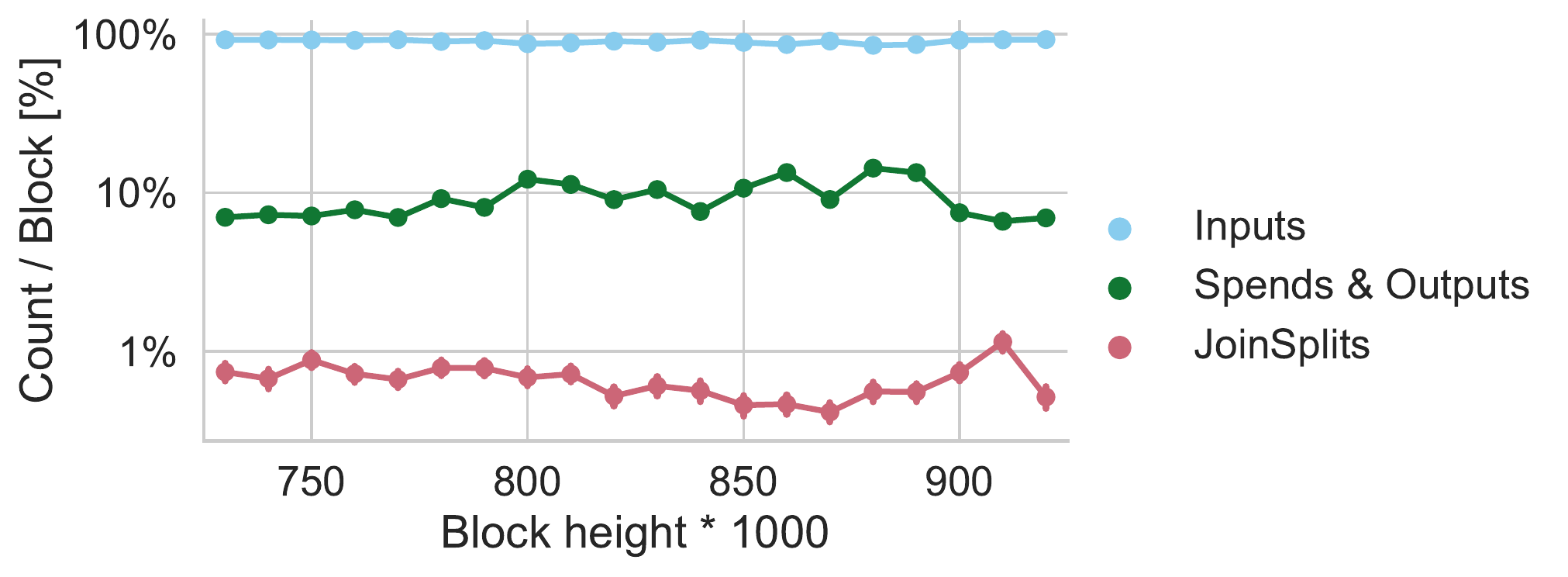}
   \caption{Ratio of transparent inputs, Spend and Output descriptions,
   and JoinSplit descriptions per block (200k blocks).}
   \label{fig:topologie}
\end{figure}

In \cref{fig:topologie}, we distinguish the different transaction types
over the last 200k blocks in our sample.
To this end, we consider a block as a set of transparent inputs,
Spend, Output, and JoinSplit descriptions.
We find that blocks consist on average of 90\,\% transparent inputs,
while Spend an Output descriptions make up 9\,\%,
and JoinSplit descriptions remain largely stable at around 1\,\% per block.

\end{document}